\documentclass[twocolumn,twocolappendix]{aastex631}

\newcommand\swift{{\it Swift}}

\begin{document}

\title{The early radio afterglow of short GRB 230217A}

\author[0000-0001-6544-8007]{G. E. Anderson}
\affiliation{International Centre for Radio Astronomy Research, Curtin University, GPO Box U1987, Perth, WA 6845, Australia}
\author[0000-0001-9915-8147]{G. Schroeder}
\affiliation{Center for Interdisciplinary Exploration and Research in Astrophysics and Department of Physics and Astronomy, Northwestern
University, 2145 Sheridan Road, Evanston, IL 60208-3112, USA}
\author[0000-0001-9149-6707]{A. J. van der Horst}
\affiliation{Department of Physics, George Washington University, 725 21st St NW, Washington, DC 20052, USA}
\author[0000-0003-2705-4941]{L. Rhodes}
\affiliation{Astrophysics, Department of Physics, University of Oxford, Keble Road, Oxford, OX1 3RH, UK}
\author[0000-0002-1195-7022]{A. Rowlinson}
\affiliation{Anton Pannekoek Institute for Astronomy, University of Amsterdam, Science Park 904, P.O. Box 94249, 1090GE Amsterdam, The Netherlands}
\affiliation{ASTRON, the Netherlands Institute for Radio Astronomy, Postbus 2, NL-7990 AA Dwingeloo, The Netherlands}
\author[0000-0003-2506-6041]{A. Bahramian}
\affiliation{International Centre for Radio Astronomy Research, Curtin University, GPO Box U1987, Perth, WA 6845, Australia}
\author[0000-0003-3507-335X]{S. I. Chastain}
\affiliation{Department of Physics and Astronomy, University of New Mexico, 210 Yale Blvd NE, Albuquerque, NM 87106, USA}
\author[0000-0002-5826-0548]{B. P. Gompertz}
\affiliation{Institute of Gravitational Wave Astronomy and School of Physics and Astronomy, University of Birmingham, Birmingham B15 2TT, UK}
\author[0000-0002-4203-2946]{P. J. Hancock}
\affiliation{International Centre for Radio Astronomy Research, Curtin University, GPO Box U1987, Perth, WA 6845, Australia}
\affiliation{Curtin Institute for Data Science, Curtin University, GPO Box U1987, Perth, WA 6845, Australia}
\author[0000-0003-1792-2338]{T. Laskar}
\affiliation{Department of Physics \& Astronomy, University of Utah, Salt Lake City, UT 84112, USA}
\author[0000-0002-9415-3766]{J. K. Leung}
\affiliation{David A. Dunlap Department of Astronomy and Astrophysics, University of Toronto, 50 St. George Street, Toronto, ON M5S 3H4, Canada}
\affiliation{Dunlap Institute for Astronomy and Astrophysics, University of Toronto, 50 St. George St., Toronto, ON M5S 3H4, Canada}
\affiliation{Racah Institute of Physics, The Hebrew University of Jerusalem, Jerusalem, 91904, Israel}
\author[0000-0002-3101-1808]{R. A. M. J. Wijers}
\affiliation{Anton Pannekoek Institute for Astronomy, University of Amsterdam, Science Park 904, P.O. Box 94249, 1090GE Amsterdam, The Netherlands}

\begin{abstract}

We present the radio afterglow of short gamma-ray burst (GRB) 230217A, which was detected less than 1 day after the gamma-ray prompt emission with the Australia Telescope Compact Array (ATCA) and the Karl G. Jansky Very Large Array (VLA). 
The ATCA rapid-response system automatically triggered an observation of GRB 230217A following its detection by the \textit{Neil Gehrels Swift Observatory} and began observing the event just 32 minutes post-burst at 5.5 and 9\,GHz for 7 hours. 
Dividing the 7-hour observation into three time-binned images allowed us to obtain radio detections with logarithmic central times of 1, 2.8 and 5.2\,hours post-burst, the first of which represents the earliest radio detection of any GRB to date. 
The decline of the light curve is consistent with reverse shock emission if the observing bands are below the spectral peak and not affected by synchrotron self-absorption. 
This makes GRB 230217A the fifth short GRB with radio detections attributed to a reverse shock at early times ($<1$\,day post-burst).
Following brightness temperature arguments, we have used our early radio detections to place the highest minimum Lorentz factor ($\Gamma_{\rm{min}}>50$ at $\sim1$\,hour) constraints on a GRB in the radio band. 
Our results demonstrate the importance of rapid radio follow-up observations with long integrations and good sensitivity for detecting the fast-evolving radio emission from short GRBs and probing their reverse shocks.

\end{abstract}

\keywords{Gamma-ray bursts --- Radio transient sources}

\section{Introduction}\label{sec:intro}

Binary neutron star (BNS) mergers are the key to understanding the behaviour of nuclear matter at extreme densities \citep{lasky14,ravi14,abbott18} and are likely the creation sites of the heaviest elements in the Universe \citep{pian17nat,levan24nat}. 
Short-duration Gamma-ray bursts (GRBs) have long been thought to arise from BNS mergers \citep{lattimer76,eichler89,narayan92,mochkovitch93}, with the near-simultaneous detection of gravitational wave (GW) event GW170817 and GRB 170817A conclusively establishing this link \citep{abbott17b}.
Short GRBs (SGRBs) are classified based on their prompt gamma-ray emission lasting $<2$\,s and being spectrally hard, while the other main class are long GRBs, which last $>2$\,s and are spectrally soft \citep[][]{kouveliotou93}. 
This emission arises from highly collimated relativistic jets driven by a black hole or rapidly spinning neutron star from the BNS merger \citep{rees92,ackermann10,rezzolla11}.  
SGRBs only represent $\sim10-20$\% of the well-localised GRB population discovered by instruments such as the
\textit{Neil Gehrels Swift Observatory} \citep[hereafter \swift;][]{gehrels04} Burst Alert Telescope \citep[BAT;][]{barthelmy05}, making them quite rare with $\lesssim10$ detected per year with \swift. 

The jet launched during an SGRB can result in a multi-wavelength afterglow as predicted by the fireball model \citep{rees92,kobayashi97,sari97,piran99,kobayashi00sari}. In this model, the relativistic ejecta interacts with the circumburst medium (CBM), producing a forward shock that accelerates electrons and generates synchrotron emission detectable from radio to X-ray (and sometimes high-energy gamma-ray) wavelengths or energies from days to years post-burst. A reverse shock is also generated, which propagates back into the shocked ejecta and can be detected as a distinct and much faster-evolving synchrotron component. This rapid evolution means that the reverse shock usually fades below detectability at X-ray and optical wavelengths within a few minutes \citep[e.g. the optical reverse shock detection $<90$\,s post-burst from the possibly short GRB 180418A;][]{becerra19}, making the radio band the most viable option for detecting and tracking its evolution \citep[e.g.][]{vanderhorst14,bright23}.

Radio observations are crucial to broadband modelling, which can map the afterglow flux density to the physical parameters of the explosion, such as its energetics and magnetic field strength as well as the density and profile of the CBM \citep{granot02}. 
Currently, only 17 SGRBs have been detected in the radio band \citep[not including GW170817;][]{berger05,soderberg06,fong14,fong15,fong21,lamb19,troja19,laskar22,anderson23GCN,rhodes23GCN,chastain24,schroeder24,Schroeder2024arXiv240713822S},
representing $\sim13\%$ of the full SGRB population with X-ray afterglows \citep{Schroeder2024arXiv240713822S}.
However, \citet{Schroeder2024arXiv240713822S} showed that if only those events detected following the VLA upgrade are considered (the instrument from which most detections are made) the detected fraction is much higher at $\sim34$\%, illustrating the importance of sensitivity for detecting SGRB radio afterglows. 

Of this sample of 17 SGRBs, 11 were detected in the radio band within $1$\,day post-burst between $4-10$\,GHz, and in several cases within a few hours. Interestingly, 8 events had faded below detectability by the time of their second or third observation between 4-9\,days post-burst. 
This indicates that $\sim50$\% of the radio-detected SGRB population may fade within a few days post-burst in the observer frame.
Meanwhile, 6 SGRBs have shown radio emission beyond 10\,days post-burst, including GRB 211106A and GRB 230205A that were detected up to $\sim100$\,days \citep{laskar22,Schroeder2024arXiv240713822S}. 
Of these, GRB 210726A is unusual as it went undetected until 11.2\,days post-burst, at which point a radio flare lasting 50\,days was detected with both MeerKAT and VLA, the likely explanation being late-time energy injection or a reverse shock from a shell collision \citep{schroeder24}.

The low number of radio-detected SGRBs may be due to the observed short radio afterglow lifetime as seen from $\sim50$\% of the detected population.
This rapid evolution indicates there should be significant variability on minute-to-hour timescales from reverse shock emission or scintillation that is likely missed by the $\sim1$\,h duration snapshot observations often obtained in the radio band \citep{fong15,Schroeder2024arXiv240713822S}.  
To ensure radio detections and comprehensive modelling of the radio afterglow, we require rapid follow-up with observations lasting for several hours post-burst. 

To increase the number of radio-detected short GRBs and probe the early-time radio parameters space, we initiated a follow-up program using the Australia Telescope Compact Array (ATCA) in 2018 that utilises the rapid-response triggering system \citep{anderson21}.  
When a short GRB is detected by \swift-BAT, the rapid-response system automatically observes the event as soon as it is above the horizon provided it is at a Declination $<+20$\,deg (with an exclusion zone of $-5<$Dec$<+5$\,deg necessary for East-West arrays). This ensures an observation occurs $<1$\,day post-burst, with integrations up to 12\,h.  
We then continue to observe the GRB three or four more times, which are logarithmically spaced in time, up to $\sim20$\,days post-burst. 
The first ATCA rapid-response observation was of short GRB 181123B, which began 12.6\,h post-burst when the source had risen above the horizon \citep{anderson21}. 
A further four events were triggered as part of this program between 2020 and 2022, three of which were observed $<1$\,day post-burst with three additional observations in the following $\sim2$ weeks, all resulting in non-detections. 
These observations were combined with quasi-simultaneous observations performed by the South African Square Kilometre Array precursor MeerKAT telescope, which enabled tight constraints to be placed on the gamma-ray efficiency and environmental density of 8 SGRBs \citep{chastain24}.

Here we present the first ATCA rapid-response radio detection of an SGRB. ATCA began observing GRB 230217A just 32 minutes post-burst for 7 hours resulting in the earliest radio detection of any GRB to date. 
Further observations with ATCA and the Karl G. Jansky Very Large Array (VLA) tracked the rapid decline of this radio emission, which faded below detectability between $5-10$\,GHz in less than $2$\,days post-burst.
In the following, we describe our observations and processing in Section 2 and radio analysis in Section 3. The GRB 230217A radio afterglow is then interpreted in the context of the SGRB population with radio detections $<1$\,day post-burst in Section 4. Throughout this paper we use the following power law representation for flux density $S \propto t^{\alpha}\nu^{\beta}$
and assume a cosmology of $H_0=67.4$ and $\Omega_m=0.315$ \citep{plank20}.

\begin{figure*}
    \centering
    \includegraphics{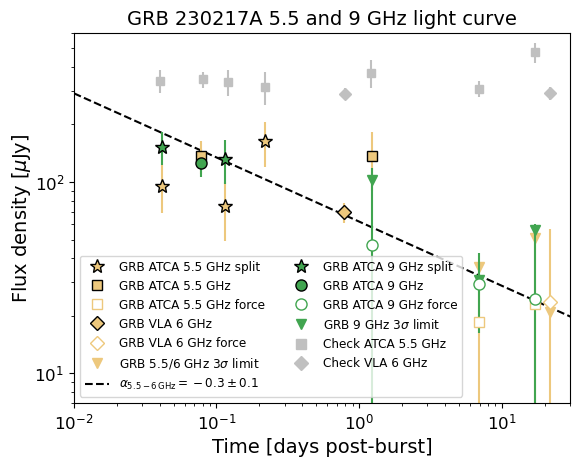}
    \caption{The radio light curve of GRB 230217A at 5.5 (ATCA observations, yellow squares/stars), 6\,GHz (VLA observations, yellow diamonds) and 9\,GHz (ATCA observations, green circles/stars).
    The stars show the afterglow detections from splitting the first ATCA observation into three and two time-binned images at 5.5 and 9\,GHz, respectively.
    The average flux density over the first ATCA observation and all subsequent ATCA observations are plotted as squares and circles at 5.5 and 9\,GHz, respectively. 
    The solid data points show detections ($>3\sigma$) whereas the open data points are force-fitted flux densities at the known position of the GRB when there was no significant detection. 
    For the non-detections, we also plot the $3\sigma$ upper limits as downward pointing triangles.
    A nearby check source detected in all ATCA 5.5\,GHz and VLA 6\,GHz images is also plotted in grey to indicate the absolute flux density scale between the ATCA and VLA measurements. 
    Over-plotted is the power law fit to the ATCA 5.5\,GHz and VLA 6\,GHz data using the average flux density from the first ATCA observation as described in Section~\ref{sec:radio_afterglow} and Table~\ref{tab:pl}. 
    }
    \label{fig:lc_data}
\end{figure*}

\section{Observations and Results}\label{sec:obs}

The short GRB 230217A was initially reported by both the \swift-BAT and the \textit{Fermi Gamma-ray Space Telescope} Gamma-ray Burst Monitor \citep{meegan09} at 21:53:10 UT on 2023 Feb 17 \citep{moss23gcn,fermi23GCN}. The \swift{} X-ray Telescope \citep[XRT;][]{burrow05} identified the X-ray afterglow \citep{capalbi23GCNa,capalbi23GCNb}. Unfortunately, the majority of XRT data was unreliable thus preventing a robust analysis of the X-ray light curve properties\footnote{https://www.swift.ac.uk/xrt\_curves/01154967/ \citep{evans07,evans09}},
which is likely due to a \swift{} Moon constraint \citep{moss23gcn} and the delay in identifying its X-ray afterglow \citep{capalbi23GCNa,capalbi23GCNb}.  
The optical afterglow was also very faint \citep{oconnor23gcb,gillanders23GCN,belles23GCN} with only a single detection reported with magnitude $I\sim24.5$ at 2.5\,d \citep{davanzo23GCN}. 
The radio afterglow was detected with both ATCA \citep{anderson23GCN} and the VLA \citep{Schroeder2024arXiv240713822S} $<1$\,day post-burst and marginally detected in a single MeerKAT observation at 5\,days \citep{chastain24}. The data processing and results of the ATCA and VLA observations are described in the following sections.

\subsection{Australia Telescope Compact Array}\label{sec:atca}

On receiving the trigger request following the \swift-BAT detection of GRB 2302017A, ATCA was on target at 2023-02-17 22:21:25 UT, just 32 minutes post-burst for 7 hours centred at the \textit{Swift}-BAT position. Follow-up observations were also obtained at 1.2, 7 and 17 days post-burst under program C3204 (PI Anderson). All ATCA observations were taken with a 6\,km maximum baseline configuration using the 4\,cm dual receiver with central frequencies of 5.5 and 9\,GHz, each with a 2\,GHz bandwidth.  
The flux and phase calibrators were PKS 1934-638 and PKS 1830-210, respectively, and the data was processed with {\sc Miriad} using standard techniques \citep{sault95}. Each image was primary beam corrected.
The flux densities were measured using the {\sc Miriad} task \texttt{imfit}. We fitted a point source at the VLA position of the GRB \citep{schroeder23GCN} to obtain a force-fitted flux density measurement for those observations with no detection. 

The radio afterglow of GRB 230217A was detected in the rapid-response observations at both 5.5 and 9\,GHz at the VLA position. As we expect the radio emission may experience variability on minute to hourly timescales $<1$\,day post-burst, we split the observed into 3 images with total integrations of 1.5, 1.5 and 4\,hours at 5.5\,GHz and 1.5 and 1.5\,hours at 9\,GHz (the last 4 hours were not imagable at 9\,GHz as the observing conditions had deteriorated). 
We also measured the average flux density from the full 7-hour image of the first ATCA observation.
The afterglow was still detected at 5.5\,GHz in a follow-up ATCA observation at 1.2\,days post-burst but had faded below detectable flux levels at 9\,GHz. All further observations were non-detections.
The $1\sigma$ flux density errors for both the GRB and check source are the \texttt{imfit} statistical error added in quadrature to a 5\% absolute flux density error, which is standard for ATCA.
The GRB flux densities (including those measured from the split and average images of the first ATCA observation), $1\sigma$ errors and image RMS near the GRB position are listed in Table~\ref{tab:flux}, with the light curves at 5.5 and 9\,GHz shown in Figure~\ref{fig:lc_data}. 

We also measured the flux density of a nearby uncatalogued radio source at RA=18:42:50.54 ($\pm0.02$\arcsec) and Dec=-28:52:02.3 ($\pm0.06$\arcsec) to test the absolute calibration between the ATCA and VLA observations. This source position was measured using the first VLA observation as it has a higher resolution than the ATCA observations (details of the VLA observations are in Section~\ref{sec:vla}). 
The 5.5\,GHz (and VLA 6\,GHz) light curve of this check source is also plotted in Figure~\ref{fig:lc_data} as grey data points showing the flux densities are consistent within $1-2\sigma$.

\subsection{Very Large Array}\label{sec:vla}

Observations with the VLA were initiated on February 18, 2023 at 16:05:18 UT for 1 hr (36\,min on target with midtime of 0.79\,days post-burst) at a mean frequency of 6\,GHz (4\,GHz bandwidth) under program 23A-296 (PI: Schroeder). Follow-up observations were obtained at 21.74 days post-burst. The flux calibrator was 3C286 and 3C48 for the first and second VLA observation, respectively, and J1820-2528 was used for the phase calibration of both observations.
The observations were performed while the VLA was in B-configuration. The VLA Science Ready Data Products produced by the Common Astronomy Software Applications \citep[\texttt{CASA};][]{mcmullin07, 2022PASP..134k4502V}  VLA Pipeline (CASA Pipeline Version 6.4.1.12) were downloaded, and the data were imaged using the VLA Imaging Pipeline\footnote{\url{https://science.nrao.edu/facilities/vla/dataprocessing/pipeline/VIPL}} (version 2023.1.0.124, \texttt{CASA} version 6.5.4). 

The flux densities of both the GRB and check source were measured from the primary beam corrected images using the {\sc Miriad} task \texttt{imfit} to be consistent with the ATCA flux measurement process. 
A 5\% absolute flux calibration error was added in quadrature to the statistical errors to produce the final $1\sigma$ flux density errors, which is standard for the VLA.
The VLA flux densities, $1\sigma$ errors that include the 5\% absolute flux calibration and the image RMS for GRB 230217A are listed in Table~\ref{tab:flux}.
The GRB and check source are plotted as yellow and grey diamonds in Figure~\ref{fig:lc_data}, respectively.
To test for short-term variability, we also split the first VLA observation into 4 images of 9 minutes each, with the resulting flux densities included in Table~\ref{tab:flux}.

\begin{table*}[]
    \centering
    \begin{tabular}{lcccccccc}
        \hline
        Telescope & Time$^{a}$ & Exposure$^{b}$ & Frequency & Flux density & $3\sigma$ threshold$^{c}$ & Frequency & Flux density & $3\sigma$ threshold$^{c}$ \\
        & (days) & (hrs) & (GHz) & ($\mu$Jy per beam) & ($\mu$Jy per beam) & (GHz) & ($\mu$Jy per beam) & ($\mu$Jy per beam) \\
        \hline
        ATCA & 0.08 & 7.1 & 5.5 & $\bf{137 \pm 27}$ & 51 & 9.0 & $\bf{125 \pm 19}$ & 39 \\ 
        & 0.04$^{\dagger}$ & 1.7 & 5.5 & $\bf{96 \pm 27}$ & 75 & 9.0 & $\bf{152 \pm 30}$ & 57 \\ 
        & 0.12$^{\dagger}$ & 1.5 & 5.5 & ${\bf 75 \pm 26}$ & 68 & 9.0 & $\bf{132 \pm 34}$ & 65 \\ 
        & 0.22$^{\dagger}$ & 4.0 & 5.5 & $\bf{163 \pm 43}$ & 86 &  &  & \\  
        & 1.22 & 3.0 & 5.5 & $\bf{136 \pm 47}$ & 109 & 9.0 & $47 \pm 72$ & 102 \\ 
        & 6.93 & 7.4 & 5.5 & $18 \pm 14$ & 36 & 9.0 & $29 \pm 13$ & 31 \\ 
        & 17.05 & 3.6 & 5.5 & $23 \pm 30$ & 51 & 9.0 & $24 \pm 36$ & 56 \\ 
        \hline
        VLA & 0.79 & 0.6 & 6.0 & $\bf{70 \pm 9}$ & 15 &  &  &  \\ 
        & 0.776$^{\ddagger}$ & 0.15 & 6.0 & $\bf{72 \pm 10}$ & 21 &  &  &  \\ 
        & 0.783$^{\ddagger}$ & 0.15 & 6.0 & $\bf{50 \pm 11}$ & 24 &  &  &  \\ 
        & 0.789$^{\ddagger}$ & 0.15 & 6.0 & $\bf{71 \pm 12}$ & 26 &  &  &  \\ 
        & 0.796$^{\ddagger}$ & 0.15 & 6.0 & $\bf{67 \pm 13}$ & 27 &  &  &  \\ 
        & 21.74 & 0.4 & 6.0 & $24 \pm 33$ & 21 &  &  &  \\
        \hline
        \hline
    \end{tabular}
    \caption{Flux density measurements of GRB 230217A. The flux densities in bold are detections. All other flux density measurements are force-fitted at the VLA position of the GRB \citep{schroeder23GCN}.\\
    $^{a}$ The central logarithmic time of the observation with respect to the \swift-BAT detection time of GRB 230217A. \\
    $^{b}$ The total exposure of the images used to derive the flux density. \\
    $^{c}$ The $3\sigma$ threshold is the image RMS multiplied by 3, which is used as an upper limit in Figure~\ref{fig:lc_data} for all non-detections.  \\
    $^{\dagger}$ Flux densities derived from splitting the first ATCA observation into three and two images at 5.5 and 9\,GHz, respectively.\\
    $^{\ddagger}$ Flux density measurements from splitting the first VLA observation into 4 images of 9 minutes each. \\
    }
    \label{tab:flux}
\end{table*}

\section{Radio afterglow analysis}\label{sec:radio_afterglow}

The radio afterglow of GRB 230217A was detected up to 1.2\,day post-burst with ATCA and VLA in the $5.5-6$\,GHz radio band. Having split the ATCA rapid-response observation into three and two images at 5.5 and 9\,GHz, respectively, the first flux density measurements has a logarithmic central time of 1\,hr post-burst (or 1.3\,hr linear central time, see stars plotted in Figure~\ref{fig:lc_data}), which is the earliest radio detection of any GRB to date.
 
We first investigate whether the flux density measurements of GRB 230217A could be affected by interstellar scintillation \citep[ISS;][]{goodman97} as has been seen in the radio afterglows of several long GRBs \citep{frail97,frail00,waxman98,chandra08,vanderhorst14,alexander19,rhodes22,anderson23}.
The all-sky model for refractive interstellar scintillation \citep[RISS19;][]{hancockRISS19} returns a transition frequency of $\nu_{0}=8.9$\,GHz, 
with a scattering screen distance of $D=1.3$\,kpc, a scattering measure $SM=9 \pm 1 \times 10^{-4}$\,kpc\,m$^{-20/3}$, and first Frenzel zone size of 2.38 $\mu$as at $\nu_{0}$.
This places our $5.5-6$\,GHz observations in the strong refractive regime and our 9\,GHz observations near the transition frequency. Following the equations in table 1 of \citet{granot14}, the predicted modulation index is up to $m=0.8$ and $m=1$ at 5.5 and 9\,GHz, respectively, for GRB 230217A.
The radio afterglow may also experience intrinsic variability due to the passage of characteristic frequencies in the afterglow emitting region(s) \citep{granot02}.

To test for evidence of short timescale variability within the first 1.2\,days post-burst, we performed a $\chi^{2}$ test on the detections of GRB 230217A, 
returning a probability of $P=0.15$ that the radio afterglow is consistent with a steady source over this time frame \citep[we require $P<0.001$ to claim variability, with $0.01<P<0.001$ being undetermined;][]{gaensler00,bell15}.
A lack of detectable variability is not surprising given that splitting the first ATCA observation reduces the sensitivity and necessarily increases the size of the error bars.
Similarly, we also perform a $\chi^{2}$ test to search for intra-observational variability within the first VLA observation when split into four $\sim9$\,min images, returning a probability of $P=0.5$ that the flux densities are consistent with a steady source within the 36 minute observation.
While scintillation may affect the GRB radio afterglow, it is not significant enough to be distinguishable in our dataset. Similarly, it is also not possible to ascertain if a characteristic frequency is evolving through the observing band within this $\sim1$\,day time frame.

The same $\chi^{2}$ test on the full $5.5-6$\,GHz GRB light curve, which includes the detections up to 1.2\,days post-burst and the force-fitted measurements for the non-detections at later times returns a probability of $P=0.0016$ of being consistent with a steady source and thus a more likely indication of variability. 
This is also compared to $P=0.12$ derived from the $5.5-6$\,GHz measurements of the check source, which is therefore steady over the 22 days of monitoring.

The constraining $3\sigma$ upper limits from the later observations (see Figure~\ref{fig:lc_data}) demonstrate the overall radio light curve trend of GRB 230217A is declining.
We fit a single power law ($S \propto t^{\alpha}$) to the $5.5-6$\,GHz and 9\,GHz light curves using \textsc{Sherpa} \citep{freeman01}. 
For both frequencies, we included the detections and force-fitted flux densities (the latter shown as open circles in Figure~\ref{fig:lc_data}). 
At both frequencies, we performed two power-law fits to the light curve: one that included a single flux density measurement over the full integration of the first ATCA observation (`Average') and the other using the flux densities measured from splitting this first observation into 3 and 2 images at 5.5 and 9\,GHz, respectively (`Split'). 
The resulting power law index and corresponding reduced $\chi^{2}$ statistic for the two $5.5-6$\,GHz fits are listed in Table~\ref{tab:pl}. The fits to the 9\,GHz light curves failed as they resulted in an unphysical reduced $\chi^{2}$ statistics near zero. 
We plot the power-law fit to the $5.5-6$\,GHz light curve that includes the average flux density from the first ATCA observation on Figure~\ref{fig:lc_data} where $\alpha = -0.3 \pm 0.1$, which we favour given its reduced $\chi^{2}\sim1$ and it is consistent with the $3\sigma$ upper limit from the final VLA observation.   
Note that the faint GRB afterglow detections meant we could not derive meaningful information on the spectral index between 5.5 and 9\,GHz. 

\begin{table}
    \centering
    \begin{tabular}{llcc}
         \hline
         $\nu$ (GHz) & First observation$^{a}$ & $\alpha$ & $\chi^{2}_{red}$  \\
         \hline
         $5.5-6$ & Split & $-0.2 \pm 0.1$ & 1.6 \\
         $5.5-6$ & Average & $-0.3 \pm 0.1$ & 0.9 \\
         \hline
    \end{tabular}
    \caption{The temporal index $\alpha$ (such that $S \propto t^{\alpha}$) and reduced $\chi^{2}$ statistic to a single power law fit to the $5.5-6$\,GHz light curve data.\\
    $^{a}$ Two fits were performed to the $5.5-6$\,GHz light curve using a single (Average) measurement or several (Split) measurements from the first ATCA observation (see Section~\ref{sec:radio_afterglow}). \\ 
    }
    \label{tab:pl}
\end{table}

\section{Discussion}

Given the limited multi-wavelength data available for GRB 230217A (see Section~\ref{sec:obs}), we restrict our investigation to just the radio afterglow. 
This includes an interpretation of the emission mechanism and a comparison to other early-time radio detections of SGRBs. 

\subsection{Reverse shock emission}

The overall trend of the 5.5 and 9\,GHz light curve of GRB 230217A is declining with a temporal index of $\alpha=-0.3 \pm 0.1$.
In Section~\ref{sec:radio_afterglow}, we demonstrate that our radio detections, which all occurred before 1.2\,days post-burst, do not provide significant evidence for scintillation, and that a characteristic frequency such as the synchrotron self-absorption break moving through the observing bands cannot be distinguished.

We now examine the detected radio emission from GRB 230217A under the reverse-forward shock framework \citep{meszaros97}. The forward and reverse shocks are both sites of particle acceleration where a population of electrons is accelerated into a power-law distribution of energies with a slope of $p$, generating two synchrotron components. The broadband spectrum of each component is described by the peak flux and three characteristic break frequencies (the synchrotron self-absorption frequency $\nu_a$, the peak frequency $\nu_m$ and the cooling frequency $\nu_c$), which evolve with time \citep{granot02}. The evolution of these breaks depends on $p$ and the slope $k$ of the density profile $\rho$ of the CBM as a function of radius $R$ ($\rho \propto R^{-k}$) where $k=0$ corresponds to a homogeneous medium and $k=2$ to a stellar wind-like medium. 
We assume that GRB 230217A is embedded within a homogeneous medium ($k=0$), which is a standard assumption for SGRBs.

Analytical solutions of the temporal power-law indices of the flux density of the radio afterglow for both the forward and reverse shock can be found in table 5 of \citet{vanderhorst14}. 
On the timescales of our radio detections of GRB 230217A (hours to $\sim1$\,day post-burst), we expect the characteristic frequencies to be ordered $\nu_a<\nu_m<\nu_c$ for both the forward and reverse shocks, and that $\nu$ is likely below $\nu_m$ and either below or above $\nu_a$ \citep{granot14}. 
For radio emission from the forward shock to decay in this same timeframe, we would require our observing frequency $\nu > \nu_m$. 
If this were the case then our temporal decay index implies $p=1.4 \pm 0.1$ for a homogeneous CBM. This is much lower than the range $2\lesssim p \lesssim 3$ expected for GRBs \citep{sari99etal}, making a forward shock interpretation unlikely.

Briefly exploring other forward shock interpretations, 
we note that the observed slope is also consistent with post-jet-break-evolution when $\nu_a<\nu<\nu_m$ (expected value of $\alpha = -1/3$). However, this implies that a jet break occurred within the first hour after the gamma-ray trigger, which would be extremely early and imply a very narrow opening angle.
We also note that the radio detections $\lesssim1$\,day post-burst are consistent with a flat light curve, which we expect for a forward shock when $\nu_a<\nu<\nu_m$ and $k=2$. However, we disfavour this interpretation as we do not expect SGRBs to be embedded in wind-like environments. 

The declining light curve better resembles reverse shock expectations, for which the $\nu_a<\nu<\nu_m$ spectral regime predicts a temporal index in the range of $-0.47\lesssim \alpha \lesssim -0.45$ \citep[depending on the thickness of the shell behind the forward shock through which the reverse shock must propagate;][]{vanderhorst14}, and is therefore consistent within $2\sigma$ of our results.
We rule out $\nu<\nu_{a,m}$ as it predicts a rising light curve. 
It is also unlikely that $\nu_a<\nu_m<\nu<\nu_c$ within the first $\sim1.2$\,days post-burst as this regime predicts much steeper temporal indices in the range of $-2.7 < \alpha < -1.3$ in a homogeneous CBM for $2 \leq p \leq 3$, which accounts for different shell thicknesses behind the forward shock.
We therefore conclude we have detected radio emission from the reverse shock of GRB 230217A and that our observing bands ($\nu$) are likely $\nu_a<\nu<\nu_m$ within 1.2\,days post-burst.

GRB 230217A was marginally detected at 1.3\,GHz with MeerKAT, measuring a flux density of $23\pm5\mu$Jy at 5\,days post-burst. This observation was the first of five taken with MeerKAT between 5-156 days post-burst \citep{chastain24}. 
Given there is no clear forward shock component in the 5.5-6 and 9\,GHz light curves, it is possible the MeerKAT detection was also of the reverse shock. 
If we interpret the rapid decline in the ATCA light curves as the reverse shock, then $\nu_{a}$ is likely already below 5.5 and 9\,GHz by the start of the first observation. 
Again following the relations
presented in table 5 of \citet{vanderhorst14}, $\nu_{a}$ evolves as $t^{-0.53}$ for a thick-shell reverse shock with $\nu_{a}<\nu_m<\nu_c$ and $k=0$ (note this temporal slope is very similar to the thin-shell case, which is between $-0.61<\alpha<-0.54$).
If we assume $\nu_a\approx9$\,GHz just before our ATCA observations at $\sim1$\,hour, we would expect $\nu_a\approx1.3$\,GHz at $\sim2$\,days.
Provided the MeerKAT frequency remained in the $\nu_a<\nu<\nu_m$ spectral regime up to 5\,days post-burst, then for an expected reverse shock decline of $\alpha \sim -0.5$, this suggests a peak flux density between $30-40\mu$Jy at 1.3\,GHz. It is therefore not unreasonable to suggest the marginal MeerKAT detection was from the fading reverse shock emission.

GRB 230217A is now the best-sampled radio reverse shock of a short GRB in the first hours post-burst (see Section~\ref{sec:comp_sgrb}), due to the 7-hour ATCA observation beginning 32 minutes post-burst that was combined with VLA and ATCA follow-up. 
However, the reduction in sensitivity caused by dividing the observation into smaller time bins (plus potential scintillation effects) makes it difficult to 
characterise the evolution of the reverse shock. 
For example, we cannot ascertain if $\nu_a$ was evolving through the observing bands during our observations or if it was already below these frequencies by the start of the first ATCA observation. 
This demonstrates that sensitivity also plays an important role (along with rapid observations with long integrations) in characterizing the reverse shock emission.

\subsection{Comparison to early-time radio detections of short GRBs}\label{sec:comp_sgrb}

\begin{figure*}
    \centering
    \includegraphics[width = 0.49\textwidth]{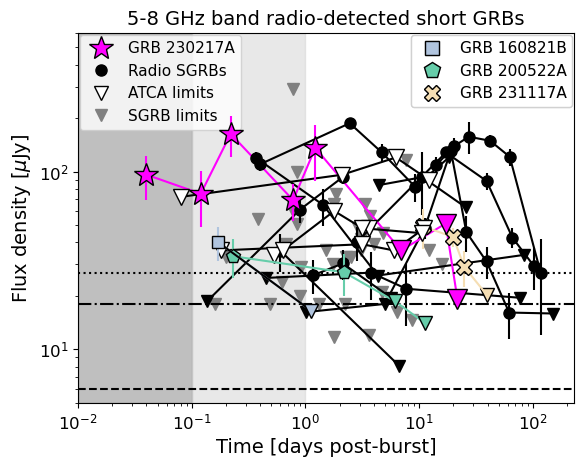}
    \includegraphics[width = 0.49\textwidth]{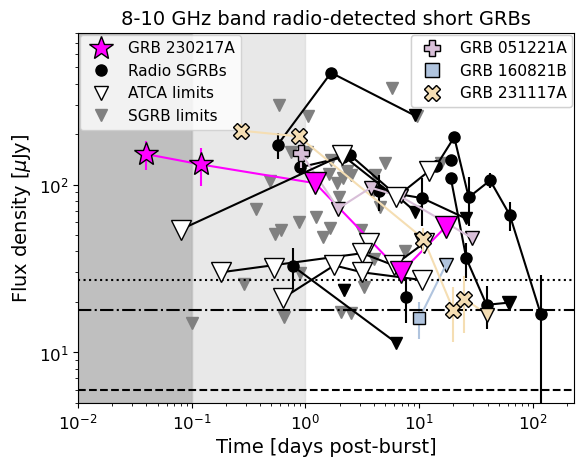}
    \caption{The light curves of SGRBs observed in the radio band between 5-8\,GHz (left) and 8-10\,GHz (right).
    The magenta stars and triangles correspond to the detections and upper limits of GRB 230217A from Table~\ref{tab:flux}.
    The other SGRBs with early time ($<1$\,day post-burst) radio emission attributed to a reverse shock from the jet hitting the CBM are also indicated with different coloured symbols \citep{soderberg06,lamb19,fong21,rhodes23GCN,Schroeder2024arXiv240713822S}.
    The black data points correspond to the other 12 radio-detected SGRBs from the literature, with circles and triangles denoting detections and upper-limits, respectively \citep{berger05,fong14,fong15,fong21,laskar22,schroeder24,Schroeder2024arXiv240713822S,levan24nat}. 
    The grey triangles show upper limits of other SGRBs not detected in the radio band taken from \citet{fong15,Schroeder2024arXiv240713822S}. 
    The white triangles are upper limits from the ATCA SGRB triggering program published in \citet{anderson21,chastain24}.
    The light grey region indicates those SGRBs with detections within 1 day post-burst and the dark grey region illustrates times $<0.1$\,days post-burst, which is a parameter space we can now probe using rapid-response observations.
    The horizontal dotted lines show the VLA $3\sigma$ sensitivity for a 9\,min image. The horizontal dashed and dot-dashed lines show the expected $3\sigma$ sensitivity of a 9\,min and 1\,min image with the Square Kilometre Array Mid, respectively (see Section~\ref{sec:conc}).
    }
    \label{fig:sgrbs}
\end{figure*}

The radio light curves of GRB 230317A are compared to all the radio-detected SGRBs with observations between $5-8$\,GHz and $8-10$\,GHz in Figure~\ref{fig:sgrbs}. 
Upper limits from observations of SGRBs not detected in these radio bands are also plotted \citep{fong15,Schroeder2024arXiv240713822S}, including the five events observed with ATCA under this rapid-response program \citep[C3204;][]{anderson21,chastain24}. Figure~\ref{fig:sgrbs} showcases the ATCA rapid-response mode's ability to probe timescales $<0.1$\,days post-burst, a previously unexplored regime, with GRB 230217A setting a new record in the earliest radio detection of a GRB.

As mentioned in Section~\ref{sec:intro}, 11 of the radio bright SGRBs ($\sim65$\%) were detected $<1$\,day post-burst, with the majority fading below detectability within a week. Of these 11 SGRBs, multi-wavelength modelling has attributed the $<1$\,day radio emission of four events to the reverse shock \citep[GRBs 051221A, 160821B, 200522A, 231117A;][]{soderberg06,lamb19,troja19,fong21,Schroeder2024arXiv240713822S}, but in each case only 1 or 2 data points sampled this synchrotron component. The light curves for these four GRBs are also highlighted in Figure~\ref{fig:sgrbs} for direct comparison with GRB 230217A. Of the remaining SGRBs with detected radio emission $<1$\,day radio post-burst, three are attributed to the forward shock \citep[GRBs 050724A, 130603B, 140903A;][]{berger05,fong14,fong15,troja16} whereas the other 4 are unmodelled. This makes GRB 230217A the fifth SGRB with radio emission attributed to a reverse shock from the jet hitting the CBM and has the densest time sampling of this emission component.

\subsection{Minimum Lorentz factor}\label{sec:lorenz}

Our early-time radio detections of GRB 230217A provide a unique opportunity to place the earliest radio constraints on the minimum Lorentz factor ($\Gamma_{\rm{min}}$) of the outflow using brightness temperature arguments \citep{galama99}. 
Usually, minimum Lorentz factors are measured from the gamma-ray emission emitted during the prompt phase of the outburst with $\Gamma_{\rm{min}} > 100-1000$ \citep{ackermann10}.
However, some important constraints have also been placed from rapid detections at other wavelengths \citep{galama99}. 
Due to the inverse squared dependence of brightness temperature on the observing frequency, early radio detections can place comparison constraints on $\Gamma_{\rm{min}}$ derived from the opposite end of the spectrum.
This type of analysis has been done for other (mainly long) GRBs that have radio detections $\lesssim1$\,day post-burst, with $\Gamma_{\rm{min}}\sim10-30$ \citep{anderson14,anderson18}. 
Meanwhile, the most constraining minimum Lorentz factors of $\Gamma_{\rm{min}}\sim20-35$ were obtained from tracking the reverse shock evolution of the luminous long GRB 221009A that allowed for an equipartition analysis \citep{bright23}.

\begin{figure}
    \centering
    \includegraphics[width=0.49\textwidth]{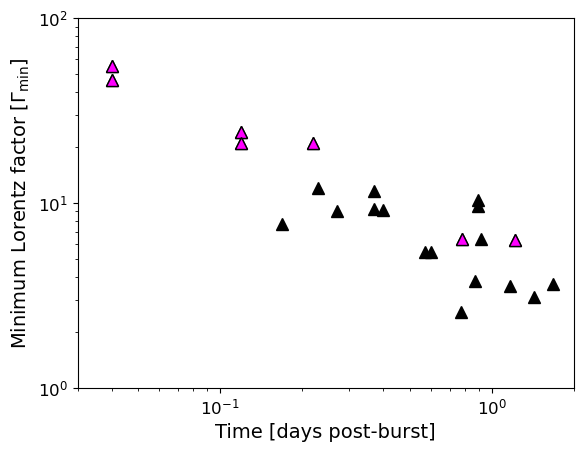}
    \caption{Lower limits on the Minimum Lorentz factors probed by early-time ($<2$\,day post-burst) radio detections of SGRBs as a function of time post-burst \citep{berger05,soderberg06,fong14,fong15,fong21,Schroeder2024arXiv240713822S,rhodes23GCN}. GRB 230217A is plotted in magenta showing the most constraining limits obtained for SGRBs in the radio band.}
    \label{fig:gamma}
\end{figure}

We calculated the brightness temperature $T_{b}$ of GRB 230217A for all detections using equation 1 from \citet{anderson14}, which assumes a non-relativistic flow such that the emission region size is simply the observing time $t$ multiplied by the speed of light. We then derived $\Gamma_{\rm{min}}$ based on the rest frame maximum brightness temperature in the inverse-Compton limit where $T_{B}\approx 10^{12}$\,K such that $T_b/T_B=\Gamma^3$ \citep{galama99}. Assuming an average radio-detected SGRB redshift of $z=0.5$ for GRB 230217A \citep{Schroeder2024arXiv240713822S}, we derived a minimum Lorentz factor of $\Gamma_{\rm{min}}>50$ and $\Gamma_{\rm{min}}>40$ from the earliest detection at $\sim1$\,hr post-burst at 5.5 and 9\,GHz, respectively. 
We also derived $\Gamma_{\rm{min}}$ for the radio SGRB population for all detections $<2$\,days post-burst, which are plotted as a function of time in Figure~\ref{fig:gamma}. These are directly compared to $\Gamma_{\rm{min}}$ from the ATCA and VLA detections of GRB 230217A, which are now the most constraining GRB limits derived from the radio band.

\section{Conclusions}\label{sec:conc}

This analysis of GRB 230217A highlights the impact of the ATCA rapid-response program for GRB studies as it probes a previously unexplored radio parameter space of $<0.1$\,days and tracks the early afterglow evolution. 
Using the ATCA rapid-response mode, we observed the short GRB 230217A just 32 minutes post-burst, obtaining the earliest radio detection of a GRB to date. Combined ATCA and VLA observations showed a fading radio light curve consistent with reverse shock emission. This makes GRB 230217A the fifth SGRB with radio detections attributed to a reverse shock at early times ($<1$\,day post-burst).
Using brightness temperature arguments, we have also placed the tightest constraint on the minimum Lorentz factor in the radio band of $\Gamma_{\rm{min}}>50$, assuming an average radio-detected SGRB redshift of 0.5.

While rapid follow-up with long integrations is key for detecting SGRB radio reverse shocks, we have not tracked their frequency-dependent evolution, which is necessary for probing the thickness of the ejecta material, the density profile of the CBM, and the energetics of the outburst, as has been possible for a small number of long GRBs \citep[e.g.][]{bright23}.  
This ATCA program already demonstrates the advantage of sensitivity and several-hour integrations over the previous GRB rapid-response program \citep{staley13,anderson18} performed with the Arcminute Microkelvin Array Large Array \citep{zwart08}, which pioneered this functionality for GRB science.
These results encourage installing similar rapid-response programs on other sensitive radio telescopes.
For example, a single 36\,min VLA observation can reach an RMS of $\sim5\mu$Jy/beam (see Table~\ref{tab:flux}), which was successfully split into four 9\,min images (see Section~\ref{sec:radio_afterglow}) with an RMS of $\sim9\mu$Jy/beam (see $3\sigma$ limit drawn as dotted lines in Figure~\ref{fig:sgrbs}, which is expected to be similar for the C(4.0-8.0\,GHz) and X(8.0-12.0\,GHz) bands in the B array configuration\footnote{https://obs.vla.nrao.edu/ect/}). 
Square Kilometre Array Mid is expected to reach an RMS of $\sim2\mu$Jy/beam and $\sim6\mu$Jy/beam at the position of GRB 230217A in a 9\,min and 1\,min observation, respectively, assuming an elevation of 45\,deg and a Briggs weighting scheme with Robust set to 0 in both the 4.6-8.5\,GHz and 8.3-15.4\,GHz bands (see $3\sigma$ limits drawn as dashed and dot-dashed lines in Figure~\ref{fig:sgrbs}).\footnote{https://sensitivity-calculator.skao.int/mid} 
A rapid-response trigger within 12\,hr post-burst that lasts for $\gtrsim3$\,hrs with either instrument would allow us to monitor the rapidly varying early-time radio emission on minute timescales with competitive sensitivity for deep and high cadence tracking of short and long GRB reverse shocks. 

\begin{acknowledgments}

We acknowledge the Gomeroi people as the traditional owners of the Australia Telescope Compact Array (ATCA) observatory site. 
The ATCA is part of the Australia Telescope National Facility, which is funded by the Australian Government for operation as a National Facility managed by CSIRO. 
We acknowledge the Whadjuk Nyungar people as the traditional owners of the land on which the Bentley Curtin University campus is located, where the majority of this research was conducted.  
We thank the referee for their recommendations on improving this manuscript.
GEA is the recipient of an Australian Research Council Discovery Early Career Researcher Award (project number DE180100346) funded by the Australian Government.
This research makes use of {\sc Astropy}, a community-developed core Python package for Astronomy \citep{TheAstropyCollaboration2013,TheAstropyCollaboration2018}, {\sc numpy} \citep{vanderWalt_numpy_2011} and {\sc scipy} \citep{Jones_scipy_2001} python modules. This research also makes use of {\sc matplotlib} \citep{hunter07}. 
This research has made use of NASA's Astrophysics Data System. 
This research has made use of SAOImage DS9, developed by the Smithsonian Astrophysical Observatory.

\end{acknowledgments}

\vspace{5mm}

\bibliography{papers}{}
\bibliographystyle{aasjournal}

\end{document}